\documentclass[runningheads,openany]{svmult}

\usepackage{makeidx}   % allows index generation
\usepackage{graphicx}  % standard LaTeX graphics tool
\usepackage{subeqnar}  % subnumbers individual equations
\usepackage{multicol}  % used for the two-column index
\usepackage{physprbb}  % modified textarea for proceedings,
\makeindex             % used for the subject index
\usepackage{amssymb}
\usepackage{epsfig}
\begin{document}

%%%%%%%%%%%%%%%%%%%%%%%%%%%%%%%%%%%%%%%%%%%%%%%%%%%%%%%%%%%%%

\renewcommand{\contriblistname}{List of Participants Contributed to the Book}
\def\lbar{\lambda\hskip-5pt\vrule height5.0pt depth-4.6pt width5pt}
\def\fr#1#2{{\textstyle{#1\over#2}}}
\hyphenation{over-determined}

%%%%%%%%%%%%%%%%%%%%%%%%%%%%%%%%%%%%%%%%%%%%%%%%%%%%%%%%%%%%%

%%%%%%%%%%%%%%%%%%%%%%%%%%%%%%%%%%%%%%%%%%%%%%%%%%%%%%%%%%%%%

{%
\author{Savely G. Karshenboim}
\authorrunning{Savely G. Karshenboim}
% if there are more than two authors,
% please abbreviate author list for running head
%
%
\institute{Max-Planck-Institut f\"ur Quantenoptik, 85748 Garching, Germany\\
%\and
D. I. Mendeleev Institute for Metrology (VNIIM), St. Petersburg 198005, Russia}

\title*{Simple Atoms, Quantum Electrodynamics\protect\newline
and Fundamental Constants}
\toctitle{Simple Atoms, Quantum Electrodynamics\protect\newline and Fundamental Constants}
\titlerunning{Simple Atoms, QED and Fundamental Constants}

\maketitle              % typesets the title of the contribution
\label{karsh}

\begin{abstract}
This review is devoted to precision physics of simple atoms. The atoms
can essentially be described in the framework of quantum electrodynamics
(QED), however, the energy levels are also affected by the effects of
the strong interaction due to the nuclear structure. We pay special
attention to QED tests based on studies of simple atoms and consider
the influence of nuclear structure on energy levels. Each calculation
requires some values of relevant fundamental constants. We discuss the accurate 
determination of the constants such as the Rydberg
constant, the fine structure constant and masses of electron, proton and
muon etc.

\end{abstract}
\index{Simple~atoms}
\index{QED|(}

\section{Introduction}

Simple atoms offer an opportunity for high accuracy calculations
within the framework of quantum electrodynamics (QED) of bound
states. Such atoms also possess a simple spectrum and some of their
transitions can be measured with high precision. Twenty, thirty years
ago most of the values which are of interest for the comparison of theory
and experiment were known experimentally with a higher accuracy than
from theoretical calculations. After a significant theoretical
progress in the development of bound state QED, the situation has
reversed. A review of the theory of light hydrogen-like atoms can
be found in \cite{report}, while recent advances in experiment and
theory have been summarized in the Proceedings of the International
Conference on Precision Physics of Simple Atomic Systems (2000) \cite{book}.

Presently, most limitations for a comparison come directly or
indirectly from the experiment. Examples of a direct experimental
limitation are the $1s-2s$ transition and the $1s$ hyperfine
structure in positronium, whose values are known theoretically better
than experimentally. An indirect {\em experimental} limitation is a
limitation of the precision of a {\em theoretical} calculation when the
uncertainty of such calculation is due to the inaccuracy of fundamental
constants (e.g. of the muon-to-electron mass ratio needed to calculate
the $1s$ hyperfine interval in muonium) or of the effects of strong
interactions (like e.g. the proton structure for the Lamb shift and
$1s$ hyperfine splitting in the hydrogen atom). The knowledge of
fundamental constants and hadronic effects is limited by the experiment
and that provides experimental limitations on theory.

This is not our first brief review on simple atoms (see e.g.
\cite{icap,limits}) and to avoid any essential overlap with 
previous papers, we mainly consider here the most recent progress
in the precision physics of hydrogen-like atoms since the publication
of the Proceedings \cite{book}. In particular, we discuss
\begin{itemize}
\item Lamb shift in the hydrogen atom;
\item hyperfine structure in hydrogen, deuterium and helium ion;
\item hyperfine structure in muonium and positronium;
\item $g$ factor of a bound electron.
\end{itemize}
We consider problems related to the accuracy of QED calculations,
hadronic effects and fundamental constants.

These atomic properties are of particular interest because of their
applications beyond atomic physics. Understanding of the Lamb shift
in hydrogen is important for an accurate determination of the Rydberg
constant $Ry$ and the proton charge radius. The hyperfine structure in
hydrogen, helium-ion and positronium allows, under some conditions, to
perform an accurate test of bound state QED and in particular to
study some higher-order corrections which are also important for
calculating the muonium hyperfine interval. The latter is a source for
the determination of the fine structure constant $\alpha$ and
muon-to-electron mass ratio. The study of the $g$ factor of a bound
electron lead to the most accurate determination of the
proton-to-electron mass ratio, which is also of interest because of a
highly accurate determination of the fine structure constant.

\section{Rydberg Constant and Lamb Shift in Hydrogen}
\noindent\index{Rydberg~constant|(}\index{Lamb~shift!h1@in~hydrogen|(}%
About fifty years ago it was discovered that in contrast to the
spectrum predicted by the Dirac equation, there are some effects in
hydrogen atom which split the $2s_{1/2}$ and $2p_{1/2}$ levels. Their
splitting known as the Lamb shift (see Fig.~\ref{figHyd}) was
successfully explained by quantum electrodynamics. The QED effects lead
to a tiny shift of energy levels and for thirty years this shift was
studied by means of microwave spectroscopy (see e.g. \cite{hinds,pipkin})
measuring either directly the splitting of the $2s_{1/2}$ and
$2p_{1/2}$ levels or a bigger splitting of the $2p_{3/2}$ and
$2s_{1/2}$ levels (fine structure) where the QED effects are
responsible for approximately 10\% of the fine-structure interval.

\begin{figure}[t]
\begin{center}
\includegraphics[width=.6\textwidth]{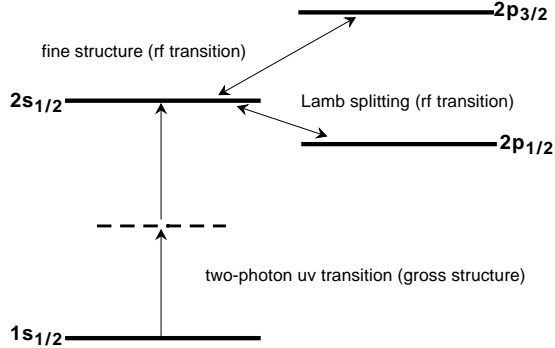}
\end{center}
\caption{Spectrum of the hydrogen atom (not to scale). The hyperfine
structure is neglected. The label {\em rf} stands for
radiofrequency intervals, while {\em uv} is for ultraviolet transitions}
\label{figHyd}
\end{figure}

The recent success of two-photon Doppler-free spectroscopy
\cite{twophot} opens another way to study QED effects directed by
high-resolution spectroscopy of gross-structure transitions. Such a
transition between energy levels with different values of the
principal quantum number $n$ is determined by the
Coulomb-Schr\"odinger formula
\begin{equation}
\label{RydLevels}
E(nl)=-\frac{(Z \alpha)^2m c^2}{2n^2}\;,
\end{equation}
where $Z$ is the nuclear charge in units of the proton charge,
$m$ is the electron mass, $c$ is the speed of light, and $\alpha$ is
the fine structure constant. For any interpretation in terms of QED
effects one has to determine a value of the Rydberg constant
\begin{equation}
\label{Ryd}
Ry = \frac{ \alpha^2m c}{2h}\;,
\end{equation}
where $h$ is the Planck constant.
Another problem in the interpretation of optical measurements of 
the hydrogen spectrum is the existence of a few levels which are significantly
affected by the QED effects. In contrast to radiofrequency
measurements, where the $2s-2p$ splitting was studied, optical
measurements have been performed with several transitions involving
$1s$, $2s$, $3s$ etc. It has to be noted that the theory of the Lamb shift for
levels with $l\neq 0$ is relatively simple, while theoretical
calculations for $s$ states lead to several serious complifications.
The problem of the involvement of few $s$ levels has been solved by introducing an auxiliary difference \cite{del1}
\begin{equation}
\Delta(n)=E_{L}(1s)-n^3E_{L}(ns)\;,
\end{equation}
for which theory is significantly simpler and more clear than for each
of the $s$ states separately.

Combining theoretical results for the difference
\cite{del2}
with measured frequencies of two or more transitions
one can extract a value
of the Rydberg constant and of the Lamb shift in the hydrogen atom. The
most recent progress in determination of the Rydberg constant is
presented in Fig.~\ref{figRy} (see \cite{twophot,codata} for
references).

\begin{figure}[bt]
\begin{center}
\includegraphics[width=.7\textwidth]{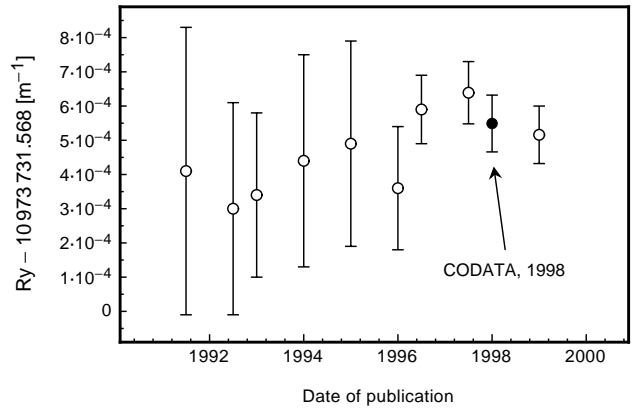}
\end{center}
\caption{Progress in the determination of the Rydberg constant by two-photon Doppler-free spectroscopy of hydrogen and deuterium. The label
{\em CODATA, 1998} stands for the recommended value of the Rydberg
constant ($Ry=10\,973\,731.568\,549(83)\;$m$^{-1}$ [10])}
\label{figRy}
\end{figure}

\begin{figure}[bt]
\begin{center}
\includegraphics[width=.7\textwidth]{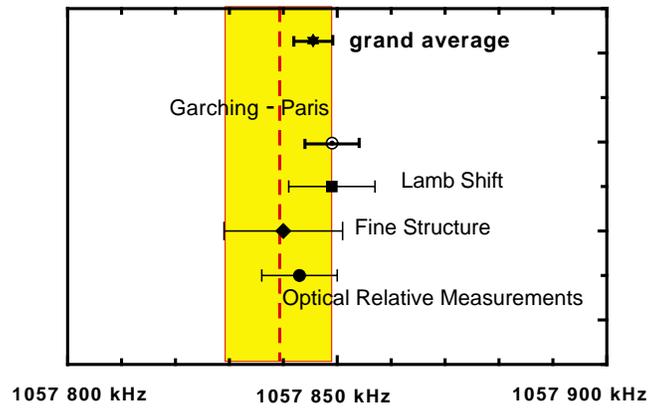}
\end{center}
\caption{Measurement of the Lamb shift in hydrogen atom.
Theory is presented according to [11]. The most accurate value
comes from comparison of the $1s-2s$ transition at MPQ (Garching) and
the $2s-ns/d$ at LKB (Paris), where $n=8,10,12$. Three results are
shown: for the average values extracted from direct Lamb shift
measurements, measurements of the fine structure and a comparison of
two optical transitions within a single experiment. The filled part is
for the theory}
\label{figLamb}
\end{figure}

\index{Nuclear~shifts~of~atomic~levels|(}\index{Proton~charge~radius|(}%
Presently the optical determination \cite{twophot,limits} of the Lamb
shift in the hydrogen atom dominates over the microwave measurements
\cite{hinds,pipkin}. The extracted value of the Lamb shift has an
uncertainty of 3~ppm. That ought to be compared with the uncertainty
of QED calculations (2~ppm) \cite{rp} and the uncertainty of the
contributions of the nuclear effects. The latter has a simple form
\begin{equation}
\label{RadP}
\Delta E_{\rm charge~radius}(nl) = \frac{2(Z \alpha)^4m c^2}{3n^3}\,\left(\frac{mcR_p}{\hbar}\right)^2\,\delta_{l0}\;.
\end{equation}
To calculate this correction one has to know the proton rms charge
radius $R_p$ with sufficient accuracy. Unfortunately, it is not known well
enough \cite{rp,icap} and leads to an uncertainty of 10~ppm for
the calculation of the Lamb shift. It is likely that a result for
$R_p$ from the electron-proton elastic scattering \cite{mainz} cannot
be improved much, but it seems to be possible to
significantly improve the accuracy of the determination of the proton
charge radius from the Lamb-shift experiment on muonic hydrogen, which
is now in progress at PSI \cite{psi}.
\index{Rydberg~constant|)}\index{Lamb~shift!h1@in~hydrogen|)}\index{Nuclear~shifts~of~atomic~levels|)}\index{Proton~charge~radius|)}

\section{Hyperfine Structure and Nuclear Effects}
\noindent\index{Nuclear~shifts~of~atomic~levels|(}%
\index{Hyperfine~structure!h1@in~hydrogen|(}%
\index{Hyperfine~structure!h2@in~deuterium|(}%
\index{Hyperfine~structure!h3@in~tritium|(}%
\index{Hyperfine~structure!he3@in~helium-3~ion|(}%
A similar problem of interference of nuclear structure and QED effects
exists for the $1s$ and $2s$ hyperfine structure in hydrogen,
deuterium, tritium and helium-3 ion. The magnitude of nuclear effects
entering theoretical calculations is at the level from 30 to 200~ppm
(depending on the atom) and their understanding is unfortunately
very poor \cite{rp,khr,d21}. We summarize the data in
Tables~\ref{17tab1} and \ref{17tabQEDHFS} (see \cite{d21}\footnote{A misprint in a value of the nuclear magnetic moment of helium-3 (it should be $\mu/\mu_B=-1.158\,740\,5$ instead of $\mu/\mu_B=-1.158\,750\,5$) has been corrected and some results on helium received minor shifts which are essentially below uncertainties} for detail).

\begin{table}
\begin{center}
\begin{tabular}{lcccc}
\hline
&&&&\\[-1.95ex]
Atom, & $E_{\rm HFS} ({\rm exp})$ & Ref. & $E_{\rm HFS} ({\rm QED})$& $\Delta E({\rm Nucl})$ \\[0.25ex]
state~~~~~~~~~~& [kHz]&&[kHz]&[ppm]\\[0.25ex]
\hline
&&&&\\[-1.95ex]
Hydrogen, $1s$ & 1\,420\,405.751\,768(1) & \protect\cite{cjp2000,exph1s} &1\,420\,452& - 33\\[0.25ex]
Deuterium, $1s$ & ~~327\,384.352\,522(2) & \protect\cite{expd1s} & ~~~327\,339&138\\[0.25ex]
Tritium, $1s$ & 1\,516\,701.470\,773(8)  &\protect\cite{mathur} & 1\,516\,760& - 36\\[0.25ex] $^3$He$^+$ ion, $1s$ & - 8\,665\,649.867(10)~~~~~~ &\protect\cite{exphe1s}  & 
- 8\,667\,494~~& - 213\\[0.25ex] \hline
&&&&\\[-1.95ex]
Hydrogen, $2s$ & 177\,556.860(15)~~~ & \protect\cite{2shydr+,2shydr} &~~~~177\,562.7  &-32\\[0.25ex]
Hydrogen, $2s$ & ~177\,556.785(29)~~~ & \protect\cite{rothery} & & - 33\\[0.25ex]
Hydrogen, $2s$ & ~177\,556.860(50)~~~ & \protect\cite{exph2s} & &- 32\\[0.25ex]
Deuterium, $2s$ & ~40\,924.439(20)~ & \protect\cite{expd2s} & ~~~~~~~40\,918.81 &137\\[0.25ex]
$^3$He$^+$ ion, $2s$ & ~- 1083\,354.980\,7(88)~~~~ & \protect\cite{prior} & 
- 1083\,585.3& - 213\\[0.25ex]
$^3$He$^+$ ion, $2s$ & ~- 1083\,354.99(20)~~~~~~~ & \protect\cite{exphe2s}  &&
- 213\\[0.25ex]
\hline
\end{tabular}
\end{center}
\caption{Hyperfine structure in light hydrogen-like atoms: QED and
nuclear contributions $\Delta E({\rm Nucl})$. The numerical results are presented for the
frequency $E/h$\label{17tab1}}
\end{table}

The leading term (so-called Fermi energy $E_F$) is a result of the
nonrelativistic interaction of the Dirac magnetic moment of electron
with the actual nuclear magnetic moment.
The leading QED contribution is related to the anomalous magnetic moment
and simply rescales the result ($E_F \to E_F\cdot(1+a_e)$). The result of the QED calculations presented 
in Table~\ref{17tab1} is of the form
\begin{equation}
E_{\rm HFS} ({\rm QED})=E_F\cdot(1+a_e)+\Delta E({\rm QED})\;,
\end{equation}
where the last term which arises from bound-state
QED effects for the $1s$ state is
given by
\begin{eqnarray}\label{QED1s}
\Delta E_{1s}({\rm QED})=
E_F&\times
&\left\{\frac{3}{2}(Z\alpha)^2+ \alpha(Z\alpha)\left(\ln2-\frac{5}{2}\right)\right.\nonumber\\
&+&{\alpha (Z \alpha )^2\over \pi}\left[-\frac{2}{3}\ln{1\over(Z\alpha)^2}
\left(\ln{\frac{1}{(Z\alpha)^2}}\right.\right.\nonumber \\
&+&\left.4\ln2-\frac{281}{240}\right) +17.122\,339\ldots \nonumber \\
&-&\left.\left.\frac{8}{15}\ln{2}+\frac{34}{225}\right]+0.7718(4)\,\frac{\alpha^2(Z\alpha)}{\pi}\right\}\,.
\end{eqnarray}
This term is in fact smaller than the nuclear corrections
as it is shown in Table~\ref{17tabQEDHFS}
(see \cite{d21} for detail). A result for the $2s$ state is of the same form with slightly 
different coeffitients \cite{d21}.

\begin{table}%[bt]
\begin{center}
\begin{tabular}{lcc}
\hline
&&\\[-2.0ex]
Atom & $\Delta E({\rm QED})$& $\Delta E({\rm Nucl})$ \\[0.2ex]
&[ppm]&[ppm]\\[0.2ex]
\hline
&&\\[-2.0ex]
Hydrogen& 23 & - 33\\[0.2ex]
Deuterium& 23&138\\[0.2ex]
Tritium & 23 & - 36\\[0.2ex]
$^3$He$^+$ ion& 108 & - 213\\[0.2ex]
\hline
\end{tabular}
\end{center}
\caption{Comparison of bound QED and nuclear corrections to the
$1s$ hyperfine interval. The QED term $\Delta E({\rm QED})$ contains
only bound-state corrections and the contribution of the anomalous
magnetic moment of electron is excluded. The nuclear contribution
$\Delta E({\rm Nucl})$  has been found via comparison of experimental
results with pure QED values (see Table 1)\label{17tabQEDHFS}}
\end{table}

From Table~\ref{17tab1} one can learn that in relative units the
effects of nuclear structure are about the same for the $1s$ and
$2s$ intervals (33~ppm for hydrogen, 138~ppm for deuterium and 213~ppm
for helium-3 ion). A reason for that is the factorized form of the
nuclear contributions  in leading approximation (cf.~(\ref{RadP}))
\begin{equation}\label{NuclPsi}
\Delta E({\rm Nucl}) = A({\rm Nucl}) \times \big\vert\Psi_{nl}
({\bf r}=0)\big\vert^2\;
\end{equation}
i.e. a product of the nuclear-structure parameter $A({\rm Nucl})$ and
a the wave function at the origin
\begin{equation}\label{psi0}
\big\vert\Psi_{nl}({\bf r}=0)\big\vert^2 = \frac{1}{\pi}
\left(\frac{(Z\alpha)m_R c}{n\hbar}\right)^3\delta_{l0}\;,
\end{equation}
which is a result of a pure
atomic problem (a nonrelativistic electron bound by the Coulomb field).
The nuclear parameter $A({\rm Nucl})$ depends on the nucleus
(proton, deutron {\em etc}.) and effect (hyperfine structure, Lamb
shift) under study, but does not depend on the atomic state.

Two parameters can be changed in the wave function:
\begin{itemize}
\item the principle quantum number $n=1,2$ for the $1s$ and $2s$ states;
\item the reduced mass of a bound particle for conventional
(electronic) atoms ($m_R\simeq m_e$) and muonic atoms ($m_R\simeq m_\mu$).
\end{itemize}
The latter option was mentioned when considering determination of the
proton charge radius via the measurement of the Lamb shift in muonic
hydrogen \cite{psi}. In the next section we consider the former
option, comparison of the $1s$ and $2s$ hyperfine interval in
hydrogen, deuterium and ion $^3$He$^+$.
\noindent\index{Nuclear~shifts~of~atomic~levels|)}\index{Hyperfine~structure!h3@in~tritium|)}%

\section{Hyperfine Structure of the $2s$ State in Hydrogen, Deuterium
and Helium-3 Ion}
\noindent\index{Hyperfine~structure!xd21@$D_{21}$|(}%
Our consideration of the $2s$ hyperfine interval is based on a study
of the specific difference
\begin{equation}
\label{d21diff}
D_{21}=8\cdot E_{\rm HFS}(2s)-E_{\rm HFS}(1s)\;,
\end{equation}
where any contribution which has a form of (\ref{NuclPsi}) should vanish.

\begin{table}%[bt]
\begin{center}
\begin{tabular}{lccc}
\hline
&&&\\[-1.9ex]
Contribution & Hydrogen & Deuterium & $^3$He$^+$ ion\\[0.3ex]
& [kHz] & [kHz] & [kHz] \\[0.3ex]
\hline
&&&\\[-1.9ex]
$D_{21}({\rm QED3})$ & 48.937 &  11.305\,6 & -1\,189.252\\[0.3ex]
$D_{21}({\rm QED4})$ & {0.018(3)} & {0.004\,3(5)}  &{-1.137(53)} \\[0.3ex]
$D_{21}({\rm nucl})$ & {-0.002} & {0.002\,6(2)} & {0.317(36)}\\[0.3ex]
\hline
&&&\\[-1.9ex]
$D_{21}({\rm theo})$ & 48.953(3) &  11.312\,5(5) & -1\,190.072(63)  \\[0.3ex]
\hline
\end{tabular}
\end{center}
\caption{Theory of the specific difference
$D_{21}=8E_{\rm HFS}(2s)-E_{\rm HFS}(1s)$ in light hydrogen-like atoms
(see [15] for detail). The numerical results are presented for the
frequency $D_{21}/h$\label{17tabD21}}
\end{table}

The difference (\ref{d21diff}) has been studied theoretically in several
papers long ago \cite{zwanziger,sternheim,pmohr}. A recent study
\cite{sgkH2s} shown that some higher-order QED and nuclear
corrections have to be taken into account for a proper comparison of
theory and experiment. The theory has been substantially improved
\cite{d21,yero2001} and it is summarized in Table~\ref{17tabD21}. The
new issues here are most of the fourth-order QED contributions
($D_{21}({\rm QED4})$) of the order $\alpha(Z\alpha)^3$,
$\alpha^2(Z\alpha)^4$, $\alpha(Z\alpha)^2{m/M}$ and $(Z\alpha)^3{m/M}$
(all are in units of the $1s$ hyperfine interval) and nuclear
corrections ($D_{21}({\rm nucl})$). The QED corrections up to the
third order ($D_{21}({\rm QED3})$) and the fourth-order contribution
of the order $(Z\alpha)^4$ have been known for a while
\cite{zwanziger,sternheim,pmohr,17breit}.

\begin{figure}[bt]
\begin{center}
\includegraphics[width=.55\textwidth]{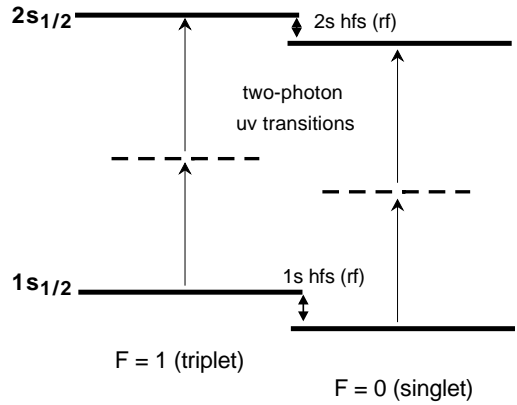}
\end{center}
\caption{Level scheme for an optical measurement of the hyperfine
structure ({\em hfs}) in the hydrogen atom (not to scale) [22]. The
label {\em rf\/} stands here for radiofrequency intervals, while {\em
uv\/} is for ultraviolet transitions}
\label{figD21Opt}
\end{figure}

For all the atoms in Table~\ref{17tabD21} the hyperfine splitting in the
ground state was measured more accurately than for the $2s$ state.
All experimental results but one were obtained by  direct measurements
of microwave transitions for the $1s$ and $2s$ hyperfine intervals.
However, the most recent result for the hydrogen atom has been
obtained by means of laser spectroscopy and measured transitions lie
in the ultraviolet range \cite{2shydr+,2shydr}. The hydrogen level scheme
is depicted in Fig.~\ref{figD21Opt}. The measured transitions were the
singlet-singlet ($F=0$) and triplet-triplet ($F=1$) two-photon $1s-2s$
ultraviolet transitions. The eventual uncertainty of the hyperfine
structure is to 6 parts in $10^{15}$ of the measured $1s-2s$
interval. The optical result in Table~\ref{17tab1} is a preliminary
one and the data analysis is still in progress.

\begin{figure}[bt]
\begin{center}
\includegraphics[width=.7\textwidth]{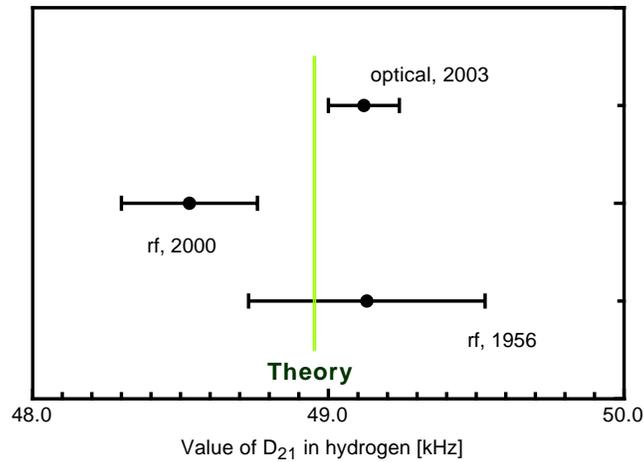}
\end{center}
\caption{Present status of measurements of $D_{21}$ in the hydrogen
atom. The results are labeled with the date of the measurement of the
$2s$ hyperfine structure. See Table~1 for references}
\label{figD21h}
\end{figure}

The comparison of theory and experiment for hydrogen and helium-3 ion
is summarized in Figs.~\ref{figD21h} and \ref{figD21he}.

\begin{figure}[bt]
\begin{center}
\includegraphics[width=.7\textwidth]{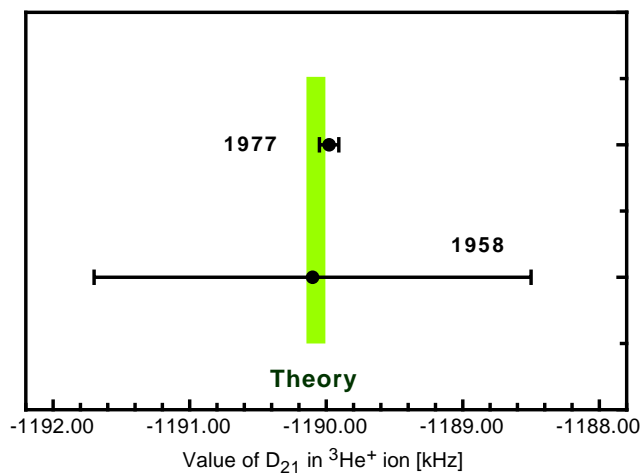}
\end{center}
\caption{Present status of measurements of $D_{21}$ in the helium ion
$^3$He$^+$. See Table~1 for references}
\label{figD21he}
\end{figure}
\noindent\index{Hyperfine~structure!xd21@$D_{21}$|)}%
\index{Hyperfine~structure!h1@in~hydrogen|)}%
\index{Hyperfine~structure!h2@in~deuterium|)}%
\index{Hyperfine~structure!he3@in~helium-3~ion|)}%

\section{Hyperfine Structure in Muonium and Positronium}

Another possibility to eliminate nuclear structure effects is
based on studies of nucleon-free atoms. Such an atomic system is to be
formed of two leptons. Two atoms of the sort have been produced and
studied for a while with high accuracy, namely, muonium and positronium.
\begin{itemize}
\item \noindent\index{Hyperfine~structure!0mu@in~muonium|(}%
Muonium is a bound system of a positive muon and electron. It can be
produced with the help of accelerators. The muon lifetime is
$2.2\cdot 10^{-6}$ sec. The most accurately measured transition is the
$1s$ hyperfine structure. The two-photon $1s-2s$ transition was also
under study. A detailed review of muonium physics can be found
in \cite{jungmann}.
\item \noindent\index{Hyperfine~structure!00p@in~positronium|(}%
Positronium can be produced at accelerators or using radioactive
positron sources. The lifetime of positronium depends on its
state. The lifetime for the $1s$ state of parapositronium (it
annihilates mainly into two photons) is $1.25\cdot 10^{-10}$ sec,
while orthopositronium in the $1s$ state has a lifetime of
$1.4\cdot 10^{-7}$ s because of three-photon decays. A list of
accurately measured quantities contains the  $1s$ hyperfine splitting,
the $1s-2s$ interval, $2s-2p$ fine structure intervals for the triplet
$1s$ state and each of the four $2p$ states, the lifetime of the $1s$
state of para- and orthopositronium and several branchings of their
decays.  A detailed review of positronium physics can be found
in \cite{conti}.
\end{itemize}

\begin{table}%[bt]
\begin{center}
\begin{tabular}{lcc}
\hline
&&\\[-1.95ex]
Term  &  Fractional~~~~ & ~~~~~~~~~~~~~~$\Delta E$~~~~~~~~~~~~~~ \\[0.25ex]
&  contribution~~ &  [kHz] \\[0.25ex]
\hline
&&\\[-1.95ex]
$E_F$ & ~~1.000\,000\,000~~~~~~ & 4.459\,031.83(50)(3)~~ \\[0.25ex]
$a_e$ & ~~0.001\,159\,652~~~~~~ & 5\,170.926(1)~\\[0.25ex]
QED2 & - 0.000\,195\,815~~~~~~ &- 873.147~~~~\\[0.25ex]
QED3 &- 0.000\,005\,923~~~~~~&- 26.410~~\\[0.25ex]
QED4 &- 0.000\,000\,123(49)&~~~~~~- 0.551(218)\\[0.25ex]
Hadronic~~~&~~0.000\,000\,054(1)~&~~~~~0.240(4)\\[0.25ex]
Weak  &- 0.000\,000\,015~~~~~~&- 0.065~~\\[0.25ex]
\hline
&&\\[-1.95ex]
Total &~~1.000\,957\,830(49)& ~~~4\,463\,302.68(51)(3)(22)\\[0.25ex]
%Exp & &4\,463\,302.68(5)\\
\hline
\end{tabular}
\end{center}
\caption{Theory of the $1s$ hyperfine splitting in muonium. The
numerical results are presented for the frequency $E/h$. The
calculations [36] have been performed for
$\alpha^{-1}= 137.035\,999\,58(52)$ [37] and
$\mu_\mu/\mu_p=3.183\,345\,17(36)$ which was obtained from the analysis of
the data on Breit-Rabi levels in muonium [38,39] (see Sect. 6)
and precession of the free muon [40]. The numerical results are
presented for the frequency $E/h$\label{17tabMu}}
\end{table}

\begin{table}%[bt]
\begin{center}
\begin{tabular}{lrr}
\hline
&&\\[-2.0ex]
Term~~~~~~  & Fractional~~~~~~ & $\Delta E$~~~~~~~~ \\[0.2ex]
&  contribution~~~~ &  [MHz]~~~~~~ \\[0.2ex]
\hline
&&\\[-2.0ex]
$E_F$ & 1.000\,000\,0~~~~~~& 204\,386.6~~~~~\\[0.2ex]
QED1 & - 0.004\,919\,6~~~~~~&-1\,005.5~~~~~\\[0.2ex]
QED2 & 0.000\,057\,7~~~~~~&11.8~~~~~\\[0.2ex]
QED3 &- 0.000\,006\,1(22)&- 1.2(5)\\[0.2ex]
\hline
&&\\[-2.0ex]
Total &0.995\,132\,1(22) & ~~~~203\,391.7(5)\\[0.2ex]
\hline
\end{tabular}
\end{center}
\caption{Theory of the $1s$ hyperfine interval in positronium. The
numerical results are presented for the frequency $E/h$. The
calculation of the second order terms was completed in [41], the
leading logarithmic contributions were found in [42], while
next-to-leading logarithmic terms %were found 
in [43]. The uncertainty
is presented following [44]\label{17tabPs}}
\end{table}

Here we discuss only the hyperfine structure of the ground state in
muonium and positronium. The theoretical status is presented in
Tables~\ref{17tabMu} and \ref{17tabPs}. The theoretical uncertainty 
for the hyperfine interval in positronium is determined only by the inaccuracy 
of the estimation of the higher-order QED effects. The uncertainty budget 
in the case of muonium is more complicated. The biggest source is
the calculation of the Fermi energy, the accuracy of which is limited by
the knowledge of the muon magnetic moment or muon mass. It is
essentially the same because the $g$ factor of the free muon is known
well enough \cite{redin}. The uncertainty related to QED is determined
by the fourth-order corrections for muonium ($\Delta E(QED4)$) and the
third-order corrections for positronium ($\Delta E(QED3)$).
These corrections are related to essentially the same diagrams (as
well as the $D_{21}(QED4)$ contribution in the previous section).
The muonium uncertainty is due to the calculation of the recoil
corrections of the order of $\alpha(Z\alpha)^2m/M$ \cite{log1,log2}
and $(Z\alpha)^3m/M$, which are related to the third-order
contributions \cite{log1} for positronium since $m=M$.

The muonium calculation is not completely free of hadronic
contributions. They are discussed in detail in
\cite{hamu1,hamu2,hamu3} and their calculation is summarized in
Fig.~\ref{figMu}. They are small enough but their understanding is
very important because of the intensive muon sources expected in future
\cite{sources} which might allow to increase dramatically the accuracy
of muonium experiments.

\begin{figure}[t]
\begin{center}
\includegraphics[width=.65\textwidth]{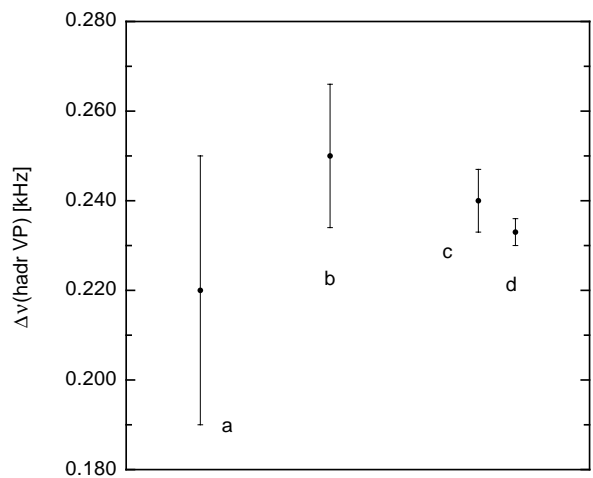}
\end{center}
\caption{Hadronic contributions to HFS in muonium. The results are
taken: $a$ from [50], $b$ from [51], $c$ from [52] and $d$ from [36,47]}
\label{figMu}
\end{figure}

A comparison of theory versus experiment for muonium is presented in
the summary of this paper. Present experimental data for positronium
together with the theoretical result are depicted in Fig.~\ref{figPo}.

\begin{figure}[t]
\begin{center}
\includegraphics[width=.5\textwidth]{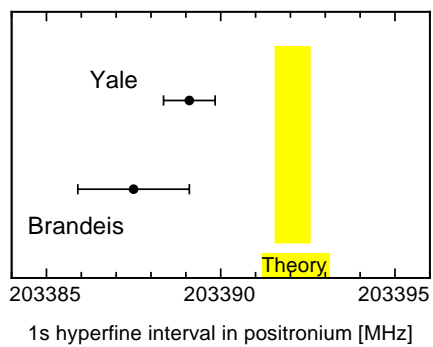}
\end{center}
\caption{Positronium hyperfine structure. The Yale experiment was
performed in 1984 [53] and the Brandeis one in 1975 [54]}
\label{figPo}
\end{figure}
\noindent\index{Hyperfine~structure!0mu@in~muonium|)}%
\noindent\index{Hyperfine~structure!00p@in~positronium|)}%

\section{$g$ Factor of Bound Electron and Muon in Muonium}
\noindent\index{g~factor!of~electron!00muonium@in~muonium|(}%
\noindent\index{g~factor!of~muon!00muonium@in~muonium|(}%
Not only the spectrum of simple atoms can be studied with high
accuracy. Other quantities are accessible to high precision
measurements as well among them the atomic magnetic moment. The interaction of an
atom with a weak homogeneous magnetic field can be expressed in terms
of an effective Hamiltonian. For muonium such a Hamiltonian has the form
\begin{equation}\label{Heff}
{\cal H} = {e\hbar\over 2 m_e} g^\prime_e \,\bigl( {\bf s}_e \cdot {\bf B} \bigr)
-{e\hbar\over 2 m_N} g^\prime_\mu \,\bigl( {\bf s}_\mu \cdot {\bf B} \bigr)
+\Delta E_{\rm HFS}\,\bigl( {\bf s}_e \cdot {\bf s}_\mu\bigr)\;,
\end{equation}
where ${\bf s}_{e(\mu)}$ stands for spin of electron (muon),
and $g^\prime_{e(\mu)}$ for the $g$ factor of a bound electron (muon)
in the muonium atom. The bound $g$ factors are now known up to the
fourth-order corrections \cite{zee} including the term of the
order $\alpha^4$, $\alpha^3m_e/m_\mu$ and $\alpha^2m_e/m_\mu$ and thus
the relative uncertainty is essentially better than $10^{-8}$. In
particular, the result for the bound muon $g$ factor reads \cite{zee}\footnote{A misprint for the $\alpha^2(Z\alpha)m_e/m_\mu$ in \cite{zee} term is corrected here}
\begin{eqnarray}
g_\mu^\prime=
g_\mu^{(0)}\cdot \biggl\{
&1&-{\alpha(Z\alpha)\over3} \left[1-{3\over2}{m_e\over m_\mu}\right]\nonumber\\
&-&{\alpha(Z\alpha)(1+Z)\over2}\left({m_e\over m_\mu}\right)^2+
{\alpha^2(Z\alpha)\over{12\pi}}{m_e\over m_\mu}
-{97\over108}\,\alpha(Z \alpha)^3\biggr\},
\end{eqnarray}
where $g_\mu^{(0)}=2\cdot(1+a_\mu)$ is the $g$ factor of a free muon.
Equation (\ref{Heff}) has been applied \cite{MuExp,MuExp1} to determine the
muon magnetic moment and muon
mass\index{Determination~of!the~muon-to-electron~mass~ratio|(} by measuring
the splitting of sublevels in the hyperfine structure of the $1s$ state in
muonium in a homogeneous magnetic field. Their dependence on the magnetic
field is given by the well known Breit-Rabi formula (see
e.g. \cite{bethe}). Since the magnetic field was calibrated via spin
precession of the proton, the muon magnetic moment was measured in
units of the proton magnetic moment, and muon-to-electron mass ratio
was derived as
\begin{equation}
\frac{m_\mu}{m_e}=\frac{\mu_\mu}{\mu_p}\,\frac{\mu_p}{\mu_B}\,\frac{1}{1+a_\mu}\;.
\end{equation}

Results on the muon mass extracted from the Breit-Rabi formula are
among the most accurate (see Fig.~\ref{figMa}). A more precise value
can  only be derived from the muonium hyperfine structure after
comparison of the experimental result with theoretical
calculations. However, the latter is of less interest, since the most
important application of the precise value of the muon-to-electron mass
is to use it as an {\em input} for calculations of the muonium
hyperfine structure while testing QED or determining the fine
structure constants $\alpha$. The adjusted CODATA result in
Fig.~\ref{figMa} was extracted from the muonium hyperfine structure
studies and in addition used some overoptimistic estimation of the
theoretical uncertainty (see \cite{hamu1} for detail).

\begin{figure}[t]
\begin{center}
\includegraphics[width=.7\textwidth]{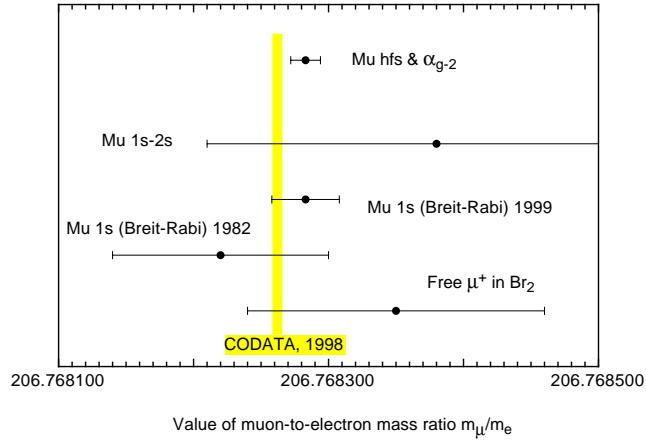}
\end{center}
\caption{The muon-to-electron mass ratio. The most accurate result
obtained from comparison of the measured hyperfine interval in
muonium [38] to the theoretical calculation [36] performed with
$\alpha_{g-2}^{-1}= 137.035\,999\,58(52)$ [37]. The results derived from
the Breit-Rabi sublevels are related to two experiments performed at
LAMPF in 1982 [39] and 1999 [38]. The others are taken from the
measurement of the $1s-2s$ interval in muonium [57], precession of a
free muon in bromine [40] and from the CODATA adjustment [10]}
\label{figMa}
\end{figure}
\noindent\index{g~factor!of~electron!00muonium@in~muonium|)}%
\noindent\index{g~factor!of~muon!00muonium@in~muonium|)}%
\noindent\index{Determination~of!the~muon-to-electron~mass~ratio|)}%

\section{$g$ Factor of a Bound Electron in a Hydrogen-Like Ion with
Spinless Nucleus}
\noindent\index{g~factor!of~electron!0ion@in~hydrogen-like~ion|(}%
In the case of an atom with a conventional nucleus (hydrogen,
deuterium etc.) another notation is used and the expression for the
Hamiltonian similar to eq.~(\ref{Heff}) can be applied. It can be used
to test QED theory as well as to determine the
electron-to-proton mass ratio. We underline that in contrast to most
other tests it is possible to do both simultaneously because of a
possibility to perform experiments with different ions.

The theoretical expression for the $g$ factor of a bound electron can
be presented in the form \cite{icap,sgkH2,pla}
\begin{equation}\label{gbound}
g^\prime_e=2\cdot\bigl(1+a_e+b\bigr)\;,
\end{equation}
where the anomalous magnetic moment of a free electron
$a_e=0.001\,159\,652\,2$ \cite{trap,codata} is known with good
enough accuracy and $b$ is the bound correction. The summary of the
calculation of the bound corrections is presented in Table~\ref{17table1}.
The uncertainty of  unknown two-loop contributions is taken from
\cite{gjetp}. The calculation of the one-loop  self-energy is
different  for different atoms. For lighter
elements (helium, beryllium), it is obtained from \cite{zee} based on
fitting data of \cite{beier}, while
for heavier ions we use the results of \cite{oneloop}. The other
results are taken from \cite{gjetp} (for the one-loop vacuum
polarization), \cite{pla} (for the nuclear correction and the electric
part of the light-by-light scattering (Wichmann-Kroll) contribution),
\cite{plb2002} (for the magnetic part of the light-by-light scattering
contribution) and \cite{recoil2} (for the recoil effects).

\begin{table}
\begin{center}
\begin{tabular}{cc}
\hline 
&\\[-1.95ex]
Ion & $g$ \\[0.25ex]
\hline
&\\[-1.8ex]
$^4$He$^+$ & 2.002\,177\,406\,7(1)\\[0.4ex]
$^{10}$Be$^{3+}$ & 2.001\,751\,574\,5(4)\\[0.4ex]
$^{12}$C$^{5+}$ & 2.001\,041\,590\,1(4)\\[0.4ex]
$^{16}$O$^{7+}$ & 2.000\,047\,020\,1(8) \\[0.4ex]
$^{18}$O$^{7+}$ & 2.000\,047\,021\,3(8) \\[0.4ex]
\hline
\end{tabular}
\end{center}
\caption{The bound electron $g$ factor in low-$Z$ hydrogen-like
ions with spinless nucleus}
\label{17table1}
\end{table}

Before comparing theory and experiment, let us shortly describe some
details of the experiment. To determine a quantity like the $g$ factor,
one needs to measure some frequency at some known magnetic field
$B$. It is clear that there is no way to directly determine magnetic
field with a high accuracy. The conventional way is to measure two
frequencies and to compare them. The frequencies measured in the
GSI-Mainz experiment \cite{werth} are the ion cyclotron frequency
\begin{equation}
{\omega_c}=\frac{{(Z-1)e}}{M_i}B
\end{equation}
and the Larmor spin precession frequency for a hydrogen-like ion with
spinless nucleus
\begin{equation}
{\omega_L}={g_b}\,\frac{{e}}{2m_e}B\;,
\end{equation}
where $M_i$ is the ion mass.

Combining them, one can obtain a result for the $g$ factor of a
bound electron
\begin{equation}\label{g2bound}
\frac{g_b}{2}={\bigl(Z-1\bigr)}\,\frac{m_e}{M_i}\,\frac{\omega_L}{\omega_c}\;
\end{equation}
or an electron-to-ion mass ratio
\begin{equation}\label{mebound}
\frac{m_e}{M_i}=\frac{1}{Z-1}\,\frac{g_b}{2}\,\frac{\omega_c}{\omega_L}\;.
\end{equation}
Today the most accurate value of ${m_e}/{M_i}$ (without using
experiments on the bound $g$ factor) is based on a measurement
of $m_e/m_p$ realized in Penning trap \cite{farnham} with a fractional
uncertainty of 2~ppm. The accuracy of measurements of $\omega_c$ and
$\omega_L$ as well as the calculation of $g_b$ (as shown in
\cite{sgkH2}) are essentially better. That means that it is preferable
to apply (\ref{mebound}) to determine the electron-to-ion mass
ratio \cite{me}.
Applying\index{g~factor!of~electron!carbon@in~hydrogen-like~carbon|(} the
theoretical value for the $g$ factor of the bound electron and using
experimental results for $\omega_c$ and $\omega_L$ in hydrogen-like
carbon \cite{werth} and some auxiliary data related to the proton and
ion masses, from \cite{codata}, we arrive at the following
values\index{Determination~of!the~proton-to-electron~mass~ratio|}
\begin{equation}\label{mpme}
\frac{m_p}{m_e}=1\,836.152\,673\,1(10)
\end{equation}
and
\begin{equation}
m_e= 0.000\,548\,579\,909\,29(31)~{\rm u}\;,
\end{equation}
which differ slightly from those in \cite{me}. The present status of
the determination of the electron-to-proton mass ratio is summarized
in Fig.~\ref{figMe}.

\begin{figure}[t]
\begin{center}
\includegraphics[width=.7\textwidth]{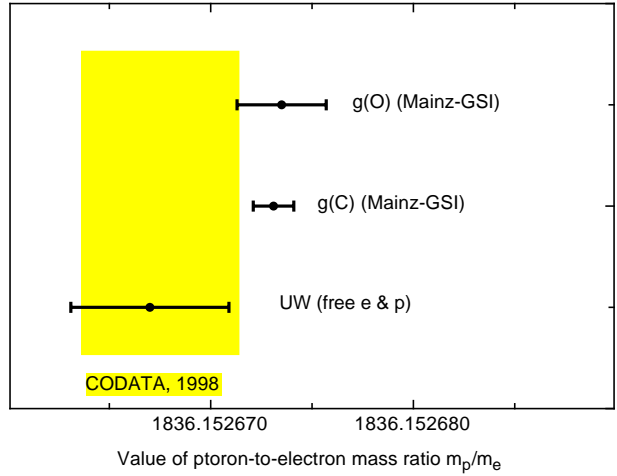}
\end{center}
\caption{The proton-to-electron mass ratio. The theory of the bound
$g$ factor is taken from Table~6, while the experimental data on the
$g$ factor in carbon and oxygen are from [68,69]. The Penning trap result
from University of Washington is from [66]}
\label{figMe}
\end{figure}

In \cite{sgkH2} it was also suggested in addition to the determination
of the electron mass to check theory by comparing the $g$ factor
for two different ions. In such a case the uncertainty related to
${m_e}/{M_i}$ in (\ref{g2bound}) vanishes. Comparing the results
for carbon \cite{werth} and
oxygen\index{g~factor!of~electron!oxygen@in~hydrogen-like~oxygen|(}
\cite{werthcjp}, we find
\begin{equation}
g(^{12}{\rm C} ^{5+})/g(^{16}{\rm O} ^{7+})
=1.000\,497\,273\,3(9)
\end{equation}
to be compared to the experimental ratio
\begin{equation}
g(^{12}{\rm C} ^{5+})/g(^{16}{\rm O} ^{7+})
=1.000\,497\,273\,1(15)\,.
\end{equation}
Theory appears to be in fair agreement with experiment. In particular,
this means  that we have a reasonable estimate of uncalculated
higher-order terms. Note, however, that for metrological applications
it is preferable to study lower $Z$ ions (hydrogen-like helium ($^4{\rm He}^{+}$)
and beryllium ($^{10}{\rm Be}^{3+}$)) to eliminate these higher-order terms.
\noindent\index{g~factor!of~electron!0ion@in~hydrogen-like~ion|)}%
\noindent\index{g~factor!of~electron!carbon@in~hydrogen-like~carbon|)}%
\noindent\index{g~factor!of~electron!oxygen@in~hydrogen-like~oxygen|)}%

\section{The Fine Structure Constant}
\noindent\index{Determination~of!the~fine~structure~constant|(}%
The fine structure constant plays a basic role in QED tests.
In atomic and particle physics there are several ways to determine
its value. The results are summarized in Fig.~\ref{figAl}. One method
based on the muonium hyperfine interval was briefly discussed in
Sect.~5. A value of the fine structure constant can also be
extracted from the neutral-helium fine structure \cite{helium,heliumexp} and
from the comparison of theory \cite{alphag2} and experiment \cite{trap}
for the anomalous magnetic moment of electron ($\alpha_{g-2}$). The
latter value has been the most accurate one for a while and there was a
long search for another competitive value. The second value
($\alpha_{Cs}$) on the list of the most precise results for the
fine structure constant is a result from recoil spectroscopy \cite{chu}.

\begin{figure}%[bt]
\begin{center}
\includegraphics[width=.7\textwidth]{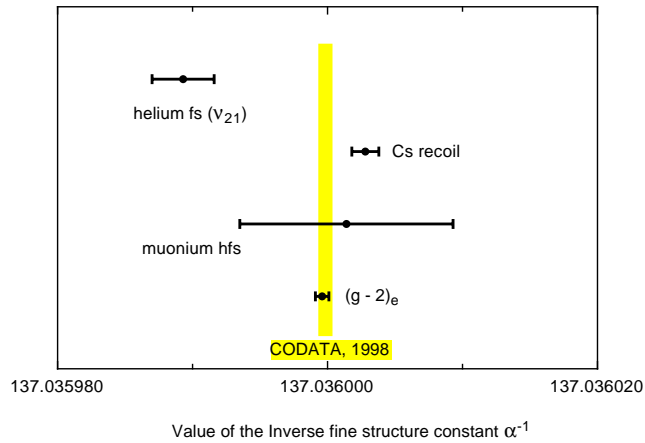}
\end{center}
\caption{The fine structure constant from atomic physics and QED}
\label{figAl}
\end{figure}

We would like to briefly consider the use and the importance of the recoil
result for the determination of the fine structure constant. Absorbing
and emitting a photon, an atom can gain some kinetic energy which can
be determined as a shift of the emitted frequency in respect to the
absorbed one ($\delta f$). A measurement of the frequency with high
accuracy is the goal of the photon recoil experiment \cite{chu}.
Combining the absorbed frequency and the shifted one, it is possible
to determine a value of atomic mass (in \cite{chu} that was caesium)
in frequency units, i.e. a value of $M_ac^2/h$. That may be compared
to the Rydberg constant $Ry=\alpha^2 m_ec/2h$. The atomic mass is
known very well in atomic units (or in units of the proton mass)
\cite{Rain}, while the determination of electron mass in proper units
is more complicated because of a different order of magnitude of the
mass. The biggest uncertainty of the recoil photon value of
$\alpha_{\rm Cs}$ comes now from the experiment \cite{chu}, while the
electron mass is the second source.

The success of $\alpha_{\rm Cs}$ determination was ascribed to
the fact that $\alpha_{g-2}$ is a QED value being derived with the
help of QED theory of the anomalous magnetic moment of electron, while
the photon recoil result is free of QED. We would like to emphasize
that the situation is not so simple and involvement of QED is not so
important. It is more important that the uncertainty of
$\alpha_{g-2}$ originates from understanding of the electron behaviour
in the Penning trap and it dominates any QED uncertainty. For this
reason, the value of $\alpha_{Cs}$  from $m_p/m_e$
in the Penning trap~\cite{farnham} obtained by the same group as the
one that determined the value of the anomalous magnetic moment of
electron \cite{trap}, can actually be correlated with $\alpha_{g-2}$.
The result
\begin{equation}
\alpha_{\rm Cs}^{-1}=137.036\,0002\,8(10)
\end{equation}
presented in Fig.~\ref{figAl} is obtained using $m_p/m_e$ from
(\ref{mpme}). The value of the proton-to-electron mass ratio found
this way is free of the problems with an electron in the Penning trap,
but some QED is involved. However, it is easy to realize that the
QED uncertainty for the $g$ factor of a bound electron and for the
anomalous magnetic moment of a free electron are very different. The
bound theory deals with simple Feynman diagrams but in Coulomb field
and in particular to improve theory of the bound $g$ factor, we need
a better understanding of Coulomb effects for ``simple'' two-loop QED
diagrams. In contrast, for the free electron no Coulomb field is
involved, but a problem arises because of the four-loop
diagrams. There is no correlation between these two calculations.
\noindent\index{Determination~of!the~fine~structure~constant|)}%

\section{Summary}

To summarize QED tests related to hyperfine structure, we present in
Table~\ref{17tabHFS} the data related to hyperfine structure of the
$1s$ state in positronium and muonium and to the $D_{21}$ value in
hydrogen, deuterium and helium-3 ion. The theory agrees with
the experiment very well.

\begin{table}%[bt]
\begin{center}\begin{tabular}{lcccc}
\hline
&&&&\\[-1.95ex]
 Atom  & ~~~~Experiment~~~~ & ~~~~~~Theory~~~~~~ &  ~~~$\Delta/\sigma$~~~ & ~~~$\sigma/E_F$~~~ \\[0.25ex]  & [kHz] & [kHz]  &   & [ppm] \\[0.25ex]  \hline
&&&&\\[-1.95ex]
Hydrogen, $D_{21}$  &  49.13(15), \protect\cite{2shydr+,2shydr} & 48.953(3)  & 1.2 & 0.10 \\[0.25ex] Hydrogen,  $D_{21}$  &  48.53(23), \protect\cite{rothery}  &  & -1.8 & 0.16 \\[0.25ex] Hydrogen,  $D_{21}$  &  49.13(40), \protect\cite{exph2s} & & 0.4 &  0.28 \\[0.25ex] Deuterium, $D_{21}$  &  11.16(16), \protect\cite{expd2s}  & 11.312\,5(5) & -1.0 & 0.49 \\[0.25ex] $^3$He$^+$ ion, $D_{21}$~~~  &-1\,189.979(71), \protect\cite{prior} &-1\,190.072(63) &1.0 &0.01  \\[0.25ex] $^3$He$^+$, $D_{21}$ & -1\,190.1(16), \protect\cite{exphe2s} &  &  0.0 & 0.18 \\[0.25ex] \hline
&&&&\\[-1.95ex]
Muonium, $1s$ & 4\,463\,302.78(5)& 4\,463\,302.88(55)& -0.18 &0.11\\[0.25ex] Positronium, $1s$ & 203\,389\,100(740) & 203\,391\,700(500) & -2.9 &4.4\\[0.25ex] Positronium, $1s$ & 203\,397\,500(1600)& & -2.5 &8.2\\[0.25ex] \hline
\end{tabular}
\end{center}
\caption{Comparison of experiment and theory of hyperfine structure in
hydrogen-like atoms. The numerical results are presented for the
frequency $E/h$. In the  $D_{21}$ case the reference is given only for
the $2s$ hyperfine interval\label{17tabHFS}}
\end{table}

The precision physics of light simple atoms provides us with an
opportunity to check higher-order effects of the perturbation theory.
The highest-order terms important for comparison of theory and
experiment are collected in Table~\ref{17tabOrd}. The uncertainty of
the $g$ factor of the bound electron in carbon and oxygen is related
to $\alpha^2(Z\alpha)^4m$ corrections in energy units, while for
calcium the crucial order is $\alpha^2(Z\alpha)^6m$.

Some of the corrections presented in Table~\ref{17tabOrd} are
completely known, some not. Many of them and in particular
$\alpha(Z\alpha)^6m^2/M^3$ and $(Z\alpha)^7m^2/M^3$ for the hyperfine
structure in muonium and helium ion, $\alpha^2(Z \alpha)^6m$ for the
Lamb shift in hydrogen and helium ion, $\alpha^7 m$ for positronium
have been known in a so-called logarithmic approximation. In other
words, only the terms with the highest power of ``big'' logarithms
(e.g. $\ln(1/Z\alpha)\sim\ln(M/m)\sim 5$ in muonium) have been
calculated. This program started for non-relativistic systems in
\cite{log1} and was developed in \cite{log2,del1,pk2,sgkH2s,d21}. By now
even some non-leading logarithmic terms have been evaluated by several
groups \cite{logps,morelogs}. It seems that we have reached some numerical
limit related to the logarithmic contribution and the calculation of the
non-logarithmic terms will be much more complicated than anything else
done before.

\begin{table}%[bt]
\begin{center}
\begin{tabular}{lc}
\hline
&\\[-1.95ex]
Value & Order \\[0.25ex]
\hline
&\\[-1.95ex]
Hydrogen, deuterium (gross structure) & $\alpha(Z\alpha)^7m$, $\alpha^2(Z\alpha)^6m$ \\[0.25ex] Hydrogen, deuterium (fine structure)              & $\alpha(Z\alpha)^7m$, $\alpha^2(Z\alpha)^6m$ \\[0.25ex] Hydrogen, deuterium (Lamb shift)              & $\alpha(Z\alpha)^7m$, $\alpha^2(Z\alpha)^6m$ \\[0.25ex] $^3$He$^+$ ion ($2s$ HFS)  & $\alpha(Z\alpha)^7m^2/M$,$\alpha(Z\alpha)^6m^3/M^2$,\\[0.25ex] & $\alpha^2(Z\alpha)^6m^2/M$, $(Z\alpha)^7m^3/M^2$\\[0.25ex] $^4$He$^+$ ion (Lamb shift)              & $\alpha(Z\alpha)^7m$, $\alpha^2(Z\alpha)^6m$ \\[0.25ex] N$^{6+}$ ion (fine structure)              & $\alpha(Z\alpha)^7m$, $\alpha^2(Z\alpha)^6m$\\[0.25ex] Muonium ($1s$ HFS)      & $(Z\alpha)^7m^3/M^2$, $\alpha(Z\alpha)^6m^3/M^2$,\\[0.25ex] &$\alpha(Z\alpha)^7m^2/M$ \\[0.25ex] Positronium ($1s$ HFS)  & $\alpha^7m$ \\[0.25ex] Positronium (gross structure)      & $\alpha^7m$ \\[0.25ex] Positronium (fine structure)       & $\alpha^7m$ \\[0.25ex] Para-positronium (decay rate)       & $\alpha^7m$ \\[0.25ex] Ortho-positronium (decay rate)      & $\alpha^8m$ \\[0.25ex] Para-positronium ($4\gamma$ branching)  & $\alpha^8m$ \\[0.25ex] Ortho-positronium ($5\gamma$ branching) & $\alpha^8m$ \\[0.25ex] \hline
\end{tabular}
\end{center}
\caption{Comparison of QED theory and experiment: crucial orders of
magnitude (see [2] for detail). Relativistic units in which $c=1$
are used in the Table\label{17tabOrd}}
\end{table}

Twenty years ago, when I joined the QED team at Mendeleev Institute
and started working on theory of simple atoms, experiment for most QED
tests was considerably better than theory. Since that time 
several groups and independent scientists from Canada, Germany, Poland, 
Russia, Sweden and USA have been working in the field and moved theory 
to a dominant position. Today we are looking forward to obtaining new 
experimental results to provide us with exciting data. 

At the moment the ball is on the experimental side and the situation 
looks as if theorists should just wait. The theoretical progress may slow 
down because of no apparent strong motivation, but that would be very 
unfortunate. It is understood that some experimental progress is possible 
in near future with the experimental accuracy surpassing the theoretical 
one. And it is clear that it is extremely difficult to improve 
precision of theory significantly and we, theorists, have to start our work on this 
improvement now. 
\index{QED|)}

\section*{Acknowledgements}
I am grateful to S. I. Eidelman, M. Fischer, T. W. H\"ansch,
E. Hessels, V. G. Ivanov, N. Kolachevsky, A. I. Milstein, P. Mohr,
V. M. Shabaev,V. A. Shelyuto, and G. Werth  for useful and stimulating
discussions. This work was supported in part by the RFBR under grants \#\# 00-02-16718, 02-02-07027, 03-02-16843.

%\newpage

\label{karsh_}
}

\end{document}